\documentclass[authoryear]{statsoc}
\usepackage[a4paper]{geometry}
\usepackage{etoolbox}

\makeatletter
\patchcmd{\@makecaption}
  {\parbox}
  {\advance\@tempdima-\fontdimen2} 
  {}{}
\makeatother  
\usepackage{graphicx}
\usepackage[textwidth=8em,textsize=small]{todonotes}
\usepackage{amsmath}
\usepackage{natbib}
\usepackage{frcursive}
\usepackage{float}
\usepackage{amsfonts}
\usepackage{enumerate}
\usepackage{textcomp}

\title[Ludometrics]{Ludometrics: Luck, and How to Measure It}
\author[]{Daniel E. Gilbert}
\address{Department of Statistical Science, Cornell University, Ithaca, NY, 14853, USA}
\email{deg257@cornell.edu}
\author{}
\author[Gilbert and Wells]{Martin T. Wells \thanks{Wells' research was partially supported by NSF-DMS 1611893 and NIH grant U19 AI111143.}}
\address{Department of Statistical Science, Cornell University, Ithaca, NY, 14853, USA}
\email{mtw1@cornell.edu}

\begin{document}
\begin{abstract}
Game theory is the study of tractable games which may be used to model more complex systems. Board games, video games and sports, however, are intractable by design, so ``ludological'' theories about these games as complex phenomena should be grounded in empiricism. A first ``ludometric'' concern is the empirical measurement of the amount of luck in various games. We argue against a narrow view of luck which includes only factors outside any player's control, and advocate for a holistic definition of luck as complementary to the variation in effective skill within a population of players. We introduce two metrics for luck in a game for a given population - one information theoretical, and one Bayesian, and discuss the estimation of these metrics using sparse, high-dimensional regression techniques. Finally, we apply these techniques to compare the amount of luck between various professional sports, between Chess and Go, and between two hobby board games: \textit{Race for the Galaxy} and \textit{Seasons}. 

\end{abstract}


\section{Introduction}
\subsection{Ludology and Ludometry}

Game theory tends to concern itself with relatively simple games like Von-Neumann matrix games, auctions, and combinatoric games with reducible complexity. These games are useful to game theorists partially because their solutions are within reach. They are useful to economists and biologists because they can be used to explain simple emergent phenomena from complex scenarios. But these games aren\textquotesingle t particularly fun. 

The games we play for fun are quite different, whether they are board games, video games, or sports. These games are outside of the realm of mathematical tractability. Board games like Chess have such expansive and irreducible game trees that the ``skill-cap'' is effectively infinite. Meanwhile, the athletic component of sports cannot be modelled by a game theorist without significant abstraction. In general, the purpose of these games is not to be solved, but to spur competition and provide an experience for the players and spectators. It is interesting, therefore, to study these games not just from the perspective of finding optimal solutions, but to understand how they foster a compelling competitive environment for necessarily sub-optimal players. The study of complex games, their mechanisms, and the experience that they foster in players has been called ``ludology''.

Studies of complex real-world phenomena ought to use empiricism to orient and evaluate models. Just as economics uses econometrics to study economies as they exist in the real world, our pursuits in ludology might be accompanied by some ``ludometrics.'' Modern online gaming platforms log an unprecedented wealth of information about who is playing which games, how the games unfold, and who wins. We can use this data to shed light on the quantitative aspects of our ludological theories. 

A pressing ludometric concern is the precise measurement of luck. Systematically quantifying luck would allow a ludologist to rigorously answer urgent questions such as ``do European board games have less luck than American board games?” or ``have video games become more luck-based over time?” First, we must agree on a definition of luck.

\subsection{Beyond Extra-Agential Luck}

A good ludometric definition of luck should be authentic and useful.  An authentic definition would capture what laymen mean when they use the word ``luck.'' A useful definition would be specific, unambiguous and facilitate the practical measurement of luck.

However, the word ``luck'' can be used in multiple, contradictory ways. An ``extra-agential'' (outside of the player) view holds luck to be ``the variation in player outcomes due to forces outside any player's control.'' By this definition, only factors such as dice, cards, spinners, weather, and random matchmaking are considered eligible sources of luck. We insist on a broader definition of luck, because an extra-agential concept of luck is neither authentic nor useful.

It isn\textquotesingle t useful because it doesn\textquotesingle t facilitate the precise measurement of luck except in the simplest of games. In most real games, the impact of dice and cards on the outcome is complex and mediated by player choices in intricate ways. It isn\textquotesingle t clear how we should account for the interactions between random events and player choice; attempting to quantify the impact of one or the other would require detailed modelling of both or else a host of simplifying assumptions. 

But more importantly, this definition isn\textquotesingle t authentic, because it places Rock Paper Scissors alongside Chess in the category of games with no luck whatsoever. While psychological exploitation yields some advantage in Rock Paper Scissors, it is commonplace for a novice to beat an experienced player, which is not the case in Chess. When we play Rock Paper Scissors, we do not feel as though we have much control over the outcome of the game, and when we win or lose, it says very little about our true Rock Paper Scissors ability, regardless of our opponent. These are the conditions under which the outcome of a game can be considered ``lucky.'' 

Here we will discuss two works which broaden the definitions of luck to include agential factors.

\subsubsection{The Success Equation}

The popular book \textit{The Success Equation: Untangling Skill and Luck in Business, Sports, and Investing} \citep{success} discusses at length the concept of luck, where it is found, and why knowledge of luck content in real-life endeavors as well as games is useful. Malboussin offers three heuristics for identifying where an activity can be placed on the ``luck-skill continuum.``
When participating in activities which are high in luck,
\begin{itemize}
    \item The outcomes are highly unpredictable
    \item Great advantages cannot be achieved through learned, repeatable behavior
    \item The amount of ``reversion to the mean'' in performance is high.
\end{itemize}

These criteria allow for a great many sources of luck which are least partially agential. These sources are categorized in the next section. 

\subsubsection{\textit{Characteristics of Games}'s Typology}

\textit{Characteristics of Games} \citep{cog} lays a groundwork for ludology with useful definitions and concepts applicable to complex games. Most pertinently, this work enumerates the potential sources of luck as follows:

\begin{itemize}
    \item 	Type I Luck or “overt randomness” is that which arises from physical randomizers such as dice, cards and spinners. 
    \item	Type II Luck is that which arises due to simultaneous decision making.
    \item	Type III Luck is that which arises due to human performance fluctuating unpredictably in complex circumstances. 
\end{itemize}

This typology more explicitly expands the definition of luck to incorporate agential factors, casting extra-agential luck as merely one type of three. Type II luck is justified by the fact competitive games with simultaneous decisions generally encourage players to act unpredictably, motivating game theorists to model player strategies as intrinsically random. Though Rock Paper Scissors is devoid of Type I Luck, the Nash Equilibrium strategy \citep{fisher} is to choose each symbol with probability $\frac{1}{3}$. Players loosely aspiring to this model will produce outcomes so arbitrary that the game is often used as a substitute for overt randomization when there are no dice in the room, hence Type II Luck. See \citet{rubinstein} for further discussion of interpreting game theoretical equilibria in real play.  

Type III Luck is more suspicious. A player's performance fluctuates from game to game; he may occasionally notice a particularly clever move in Chess, or he may trip over his feet in a game of Tennis. These anomalies can be unilaterally attributed to a player, so why should they be factored into luck? \citet{cog} concisely argues for the necessity of “Type III Luck” with the hypothetical game \textit{Guess the Digit of Pi}.  In this game, one player chooses an eight-digit number, and the other player must name that digit of $\pi$ in 10 seconds or else lose the game. It is beyond a human’s capacity to calculate the solution so quickly (or memorize so many numbers), so most of the time, the player is compelled to guess. This game is devoid of Type I or Type II Luck, and thus suggests a third type of luck emergent from players facing intractability. 

Some of this same intractability can be found in Chess's unfathomable game tree, which suggests that Chess has at least some luck. This claim can be a sticking point for those who relish the idea of a game of ``pure skill.'' However, this categorization was hardly defensible in the first place; a random number generator could beat Magnus Carlsen once every few heat-deaths of the universe, and the reason is unlikely due to the machine suddenly acquiring a high level of skill, and then immediately relinquishing it. 

There is one source of luck missing from \textit{Characteristics of Games}'s typology, which we will call

\begin{itemize} 
\item Type IV Luck: luck due to matchmaking.
\end{itemize}

Consider a population of Rock Paper Scissors players, each of whom will be matched with a random opponent for a single game of Rock Paper Scissors. Because players will only need to throw one symbol, each has chosen his move in advance. The player strategies are therefore deterministic, so whether a player wins or loses depends entirely on the random selection of his opponent. Without a concept of ``matchmaking luck,'' this game has no luck whatsoever. However, if players were to choose their moves a split second after matchmaking rather than a split second before, the game would suddenly be almost entirely luck. Thus, a non-arbitrary definition of luck must accommodate uncertainty due to matchmaking. 

\subsection{Previous Methods for Measuring Luck}

Exhaustively sourcing and categorizing luck is interesting to the game designer and ludologist, because different types of luck may result in subjectively different player experiences. But this typology does not help us quantify luck. For instance, games wherein a single die roll decides the outcome likely has more luck than a game with many die rolls which influence and depend on player choices. The total luck depends not on the number of die rolls but their influence on the outcome of the game. 

\subsubsection{\textit{Characteristics of Game's Method}}

In addition to categorizing sources of luck, \citet{cog} proposes a conception of a game's total luck content as follows. For a given two-player game, we could carefully select players from the population of players and line them up so that each would have at least a 60\% win-rate against the player to their right. We could then characterize the total skill content of a game by the longest skill chain we could construct from the population of players of that game. A game like Chess would have a long chain with 10-15 players, whereas a game like Yahtzee may only have a chain 2-3 players long, indicating that Yahtzee has far more luck than chess.
While this method would be difficult to execute, we agree that it essentially captures the amount of luck in a game (albeit in low resolution), and note the following features of this conception of luck:

\begin{itemize}
    \item It is contingent on the population of players.
    \item It does not depend on the particulars of how a given match transpires, but only on the relationship between the players and their likely match outcomes.
    \item It is counterfactual in the sense that it depends not on matches which necessarily occurred, but upon hypothetical matches against certain representative opponents from the population.
\end{itemize}

\subsubsection{\textit{The Success Equation's Method}}

\citet{success} proposes a more practical method for measuring luck with similar features. To quantify the amount of luck in various professional sports, Malboussin uses the following procedure: 

Using the schedule of matches for a season of the given sport with $T$ teams, where each team $i \in \{1,\dots,T\}$ has played $n$ games. 
\begin{enumerate}[1)]
\item Compute $p_i, i \in \{1,\dots,T\}$,  the proportion of games won for each team. 
\item Compute $V = \frac{1}{T-1}\sum_{i=1}^T (p_i - \bar{p})^2$, the sample variance of the team win-rates.
\item Set $V_0 = .25n$, the hypothetical variance of the win-rates if the game were in fact pure luck.
\item Then the amount of luck in the game is $V_0/V$.
\end{enumerate}

Thus, if the population of teams exhibits a great spread in win-rates, the game will have be considered low in luck, whereas if all of the teams have about a 50\% win-rate, the game will be considered high in luck. \citet{success} uses this method to construct an approximate luck-skill continuum of professional sports, finding that NHL seasons are relatively high in luck, whereas NBA seasons are relatively low in luck. 

This definition is population contingent and depends only on the match schedule including the outcomes. However, it is not counter-factual, in that team win-rates, which depend on the opponents each team happens to face, are used directly in the formula for luck. We view this as a weakness: some teams will happen to be matched against relatively strong opponents, and their win-rates will suffer. Thus, this method confounds sampling happenstance in match schedules with the effect of team skill. Furthermore, this method requires perfectly balanced data, and thus does not generalize well beyond professional sports. 

\subsection{Defining Luck}

Here we offer our own general concept of total luck, which elaborates on the spirit of the previous conceptions. We will then formalize our definition mathematically in Section \ref{ludodef}.

The amount of luck in a game is somehow related to a concept of skill. However, the common conception that games exist on a spectrum between ``skill-based'' and ``luck-based'' is overly simplistic. As \citet{cog} explains, games can exist with any combination of ``skill-content'' and ``luck-content'' levels. Table \ref{tab:tab1}, from \cite{cog}, illustrates that Poker and Chess are games with undiscovered skill-caps, but the results of each individual hand of Poker are quite arbitrary.

\begin{table} \label{tab:tab1}
\caption{Examples of skill-content and luck-content combinations.}
\centering
    \begin{tabular}{c|c c}
         & Low Skill & High Skill\\
         \hline
         Low Luck & Tic-Tac-Toe & Chess\\
         High Luck & Slots & Poker\\
    \end{tabular}
\end{table}

This concept of ``total skill'' is mercurial. When we refer to a player's level of skill at a particular game we refer to his propensity for playing it, whether due to advanced knowledge or tailored athleticism. A player's skill varies only slowly over time, as opposed to his performance, which may fluctuate greatly from game to game.

Furthermore, skill is multidimensional; a player may excel at one aspect of a game while remaining weak at others. Chess players have been known to be weak to certain openings and strong against others. In certain games, Player A may usually beat Player B, who usually beats Player C, who usually beats Player A.

Finally, skill cannot be determined solely from a player's game outcomes. The best Poker players are more ``skilled'' at Poker than the best Bezique players are at Bezique, even though they may achieve similar win-rates. Poker boasts a committed community of players who have won their mastery through years of study using wisdom developed for decades; it is difficult to rise to the top of the scene. Bezique, on the other hand is played casually by an obscure smattering of card game hobbyists; a moderately motivated individual could expect to become one of the best players within months or years \citep{play}.

\subsection{``Returns to Skill'' and Luck}

Thus, in order to relate luck and skill, we should make use of \textit{CoG}'s concept of ``Returns to Skill'' (RS). The returns to skill is the degree to which skill, differentially expressed within a population, determines player outcomes.  

Suppose a game has a corresponding population of players. In theory, each player has a probability distribution over the outcomes they would attain if matched with an opponent (or set of opponents) uniformly at random from the population. For instance, a chess player rated 1300 may have about a 20\% chance of winning against a random opponent, and a 2000-rated player may have a 95\% chance of winning against a random opponent. Then RS is a measure of how much these marginal player outcome distributions vary over the population. Luck is complementary; a game in which knowing the player told us little about his outcome distribution would be low in RS but high in luck. 

This definition is authentic and useful. It is authentic because it attributes a regularity in skill to players and relegates the fluctuations in player performance to luck. Moments in games which fail to reflect the regular component of a players’ skill are precisely those which are considered ``lucky,'' regardless of whether they arise from dice or entirely agential chaos. Further, this definition allows us to ascertain the amount of luck in a game using only data which identifies its players and their outcomes - the game itself may as well be a black box. The methodology for ascertaining luck is developed in the following sections.

Before doing so, we elaborate on two essential features of RS and luck.

\noindent{\it Population Contingency}\\
RS and luck in a game depend on the population. This is unavoidable. If the population of players is limited to those with very similar skill, the RS will be found to be low and the luck will be found to be high. Thus, to compare the amount of luck inherent to different games, we will need to assume that the reference populations have similar distributions of expertise.  We may also consider the amount of luck in a game with respect to a sub-population; for instance, we might find that amateur Chess is luckier than amateur Go, but professional Chess has less luck than professional Go. With some additional assumptions and modelling, we may be able to satisfy questions about the luck associated with a theoretical population, such as a ``standard population” with a certain distribution of hours of experience.

\noindent{\it Uniform Matchmaking}\\
RS and luck depend on the distribution of player outcomes under uniformly random matchmaking. We don’t usually find uniform matchmaking in the wild. Players often match themselves against similarly skilled opponents, especially in the case of competitive video games, for which sophisticated matchmaking systems ensure a near 50\% win-rate for every player. The reason we do not conclude that all of these games are pure luck is because we occasionally have the pleasure of watching a seasoned professional trounce a novice. The RS of a game is determined by what would hypothetically happen between players of different skill levels, even if they never play each other. In the absence of uniformly matched data, we must use some assumptions and modeling to extract this hypothetical information from the arbitrarily matched data we \textit{do} observe.

\section{Ludometric Definitions}
\label{ludodef}
\subsection{Discrete Outcome Games}

For the purpose of studying player outcomes, let the $n$-player version of a game $\mathcal{G}$ be represented by $\mathcal{G}_n = \{\mathbb{O},\mathbb{A},\mathbb{M}, \mathbb{P} \}$. Here,
\begin{itemize}
    \item $\mathbb{O}$ is the set of possible outcomes. 
\end{itemize}
For example, in Chess, 
    $$\mathbb{O} = \left \{ \{\text{Win, Lose}\},\{\text{Tie, Tie}\},\{\text{Lose, Win}\}\right \}$$
    In an $n$-player game of \textit{New Angeles}, where any number of players may win as long as at least one player loses, 
    $$\mathbb{O} = \{o \in \{\text{Win, Lose}\}^n \mid \sum_{i=1}^n \mathbf{1}_{\{o_i = \text{Lose}\}} \geq 1 \}$$
\begin{itemize}

\item $\mathbb{A}$ is a finite population of players (``\textbf{A}gents''). For a game with a population of $N$ players, $\mathbb{A} = \{1,2,\dots,A\} $.

    \item $\mathbb{M}$ is a ``matchmaking'' probability function with domain $\mathbb{A}_n$, the set of all combinations of $n$ players from $\mathbb{A}$.
    \item $\mathbb{P} = \left \{ P_\mathbf{A} \mid \mathbf{A} \in \mathbb{A}_n \right \}$ is a set of probability functions, each over the outcomes $\mathbb{O}$, which depend on the set of players, $\mathbf{A}$. 
    
\end{itemize}

Thus an instance of game $\mathcal{G}_n$ is a random variable $G = \{\mathbf{O} \in \mathbb{O}, \mathbf{A} \in \mathbb{A}_n\},$ where $\mathbf{A} \sim \mathbb{M}$ and $\mathbf{O}|\mathbf{A} \sim P_\mathbf{A}$.  Our formulation of $\mathcal{G}_n$ is reminiscent of the theory of games and statistical decisions framework developed in \cite{blackwell}.

Notation: We write $\mathbf{A} = (A_1,  A_{-1})$ and $\mathbf{O} = (O_1, O_{-1})$. Thus $O_1$ is the outcome associated with player $A_1$. (The order of this player vector should not be conflated with in-game seating, which should be arbitrarily assigned during the game). Let $\mathbb{O}^{(1)} = \{o_1 \mid \text{ } \exists \text{ }{o}_{-1} \text{ s.t. }(o_1,o_{-1}) \in \mathbb{O} \}$ be the space of possible outcomes for an individual player. Let $\mathbb{A}_{a_1}^{(-1)} = (\mathbb{A} \backslash a_1)_{n-1}$, the possible opponent sets contingent on $a_1$. Then let $\mathbb{O}_{o_1}^{(-1)} = \{o_{-1} | (o, o_{-1}) \in \mathbb{O} \}$ be the space of possible outcomes for the remaining players contingent on $o_1$.

For $o \in \mathbb{O}^{(1)}$ and $a \in \mathbb{A}$, let 
\begin{align*}
    P_a(o) &\equiv P(\mathbb{O}_1 = o \mid A_1 = a) \\
    &= \sum_{a_{-1} \in \mathbb{A}_{a}^{(-1)}} \sum_{o_{-1} \in \mathbb{O}_{o}^{(-1)}} \mathbb{M}(A_{-1} = a_{-1} \mid A_1 = a) P_{(a,a_{-1})}(O_1 = o) \\
    P(o) &\equiv P(O_1 = o) \\
    &= \sum_{\mathbf{a} \in \mathbb{A}_n} \sum_{o_{-1} \in \mathbb{O}_{o}^{(-1)}} \mathbb{M}(\mathbf{A} = \mathbf{a}) P_\mathbf{a}(O_1 = o)
\end{align*}

Thus $P_a$ is the distribution of outcomes for a particular player $a$ under matchmaking system $\mathbb{M}$, and $P$ is the overall population's distribution of outcomes under matchmaking system $\mathbb{M}$. If $\mathbb{M}(\mathbf{A} = \mathbf{a}) = \binom{N}{n}^{-1}$ for all $\mathbf{a} \in \mathbb{A}_n$, $\mathbb{M}$ is considered a uniform matchmaker.

Then the ``returns to skill,'' $\mathcal{S}$, in $\mathcal{G}_n$ is defined as:
\begin{align}
    \mathcal{S} &\equiv \frac{ \underset{o \in \mathbb{O}^{(1)}}{\sum} P(o) \log P(o) - \frac{1}{N} \underset{a \in \mathbb{A}}{\sum} \underset{o \in \mathbb{O}^{(1)}}{\sum} P_a(o) \log P_a(o)}{\underset{o \in \mathbb{O}}{\sum} P(o) \log P(o) } \\
    &= \frac{I(O_1,A_1)}{H(O_1)} \text{  under a uniform matchmaker.}
\end{align}
where $I(O_1,A_1)$ is the mutual information between a player and his outcome, and $H(O_1)$ is the unconditional entropy of $O_1$. Thus $\mathcal{S}$ is a measure of relative information, $0 \leq \mathcal{S} \leq 1,$ and the ``luck'' in $\mathcal{G}_n$ is:
\begin{align}
    \mathcal{L} \equiv 1 - \mathcal{S}.
\end{align}

\subsection{Continuous Outcome Games}

For some games, we might want to consider a continuous outcome space; for instance $\mathbb{O} = \mathbb{R}^n$, as in a betting game with arbitrary monetary outcomes. 

In this case, we replace our discrete probability mass functions with densities: for $o \in \mathbb{O}^{(1)}$, $\mathbb{P} = \left \{ f_{\mathbf{a}} \mid \mathbf{a} \in \mathbb{A}_n \right \} $, $f_a(o) \equiv f_{O_1|A_1}(o) $, and
 $ f(o) \equiv f_{O_1}(o) $.

Information theoretic quantities do not retain all of their properties in the continuous setting. Here, $H(O)$ may be negative, and thus the relative information is no longer a natural probabilistic representation of luck. In this case, we might define 
\begin{align}
    \mathcal{S} &= 1 - \exp \left \{\int_{o \in \mathbb{O}^{(1)}} f(o) \log f(o) \text{d}o - \frac{1}{N} \underset{a \in \mathbb{A}}{\sum} \int_{o \in \mathbb{O}^{(1)}} f_a(o) \log f_a(o) \text{d}o\right \} \\
    &= 1 - \exp \{ -I(O_1,A_1)\}  \text{  under a uniform matchmaker.}
\end{align}

Then $0 \leq \mathcal{S} \leq 1$ and we may once again define luck as $\mathcal{L} = 1 - \mathcal{S}$.

Due to the relative rarity of games with continuous outcomes, we will focus the remainder of the discussion on discrete outcome games. 

\section{Estimation}

The game outcome set $\mathbb{O}$ is inherent to the rules of a game, and we assume the population of players $\mathbb{A}$ is known. The definition of luck is contingent on a uniform matchmaker $\mathbb{M}$. Therefore, in order to quantify the amount of luck in a game, we need only estimate $P_a$ for each player $a \in \mathbb{A}$. 

If we were lucky enough to observe a large number of uniformly-made matches, we could quite simply estimate each $P_a$ using $\hat{P}_a$, the empirical distribution of each player's outcomes, unconditional on the opponents. 

However, in the ubiquitous case we observe arbitrarily matched, unbalanced data, we require some assumptions and modeling.

\subsection{Skill Models}

Let $\mathbf{s} = \{s_a \in S, a \in \mathbb{A}\}$ be a set of skill levels associated with the population of players. Usually skill levels are scalar, i.e. $S = \mathbb{R}$, but we may consider multi-dimensional skill levels, e.g. $S = \mathbb{R}^q$. We will often need to constrain the skill levels (to mean 0, for instance). Therefore, let $\mathbb{S} = \{\mathbf{s} \in S^n \mid c(\mathbf{s}) \in C\}.$

To construct a skill model for $\mathbb{P}$, we assume, for $o \in \mathbb{O}$ and possibly some parameters $\theta \in \Theta$, $\mathbb{P}_\mathbf{a}(o) = P^\theta_{\mathbf{s}_\mathbf{a}}(o)$, i.e. the outcome probabilities depend on the players only through their skill levels. 

Suppose we observe a set of arbitrary matches $(o_i, \mathbf{a}_i)$ for $i = 1,\dots,m$. Under the assumption 
\begin{align*}
    o_i \overset{i.i.d}{\sim} P^\theta_{\mathbf{s}_{\mathbf{a}_i}} \text{ for } i = 1,\dots,m
\end{align*}
a maximum likelihood rule for $\mathbb{P}$ is:
\begin{align}
    (\hat{\theta},\hat{\mathbf{s}}) = \underset{\theta \in \Theta, \mathbf{s} \in \mathbb{S}}{\text{argmax}} \prod_{i=1}^m P^\theta_{\mathbf{s}_{\mathbf{a}_i}}(o_i). \label{mle}
\end{align}

\subsubsection{Bradley-Terry Model}

The Bradley-Terry Model \citep{bradter} is a well-known model for inferences from paired comparisons.  For the Bradley-Terry model for two-player games without ties, let \\
$\mathbb{O} = \left\{\{\text{Win, Lose}\},\{\text{Lose, Win}\} \right\}$, $S = \mathbb{R}$, $\sum_{a \in \mathbb{A}} s_a = 0$, $n = 2$, and 
\begin{align}
    P_{a_1,a_2}(o) = \begin{cases}
      \frac{1}{1+e^{-(s_{a_1} - s_{a_2})}} & \text{if}\ o=(\text{Win, Lose}) \\
      \frac{1}{1+e^{(s_{a_1} - s_{a_2})}} & \text{if}\ o=(\text{Lose, Win}).
    \end{cases} \label{brad}
\end{align}

This model can be interpreted as a ``latent performance model,'' wherein players exhibit performances in each game randomly distributed about their skill levels.

Let $$Y_{a_i} \overset{\text{ind}}{\sim} EV_1(s_{a_i}, 1) \text{ for } i = 1,2 $$
be each player's unobserved random performance, where $EV_1$ is the Type-1 Extreme Value or Gumbel distribution \citep{Gumbel}. Then if 
\begin{align}
    O|Y_{a_1},Y_{a_2} = \begin{cases}
(\text{Win, Lose}) & \text{if}\ Y_{a_1} > Y_{a_2} \\
(\text{Lose, Win}) & \text{otherwise}
\end{cases} \label{2plperf}
\end{align}
it follows that (\ref{brad}) holds.

This Bradley-Terry model is equivalent to a logistic regression model with coefficient parameters $\mathbf{s}$, a canonical (logit) link function and design matrix $\mathbf{x}$, where the $i$th row of $\mathbf{x}$, $x_{i \cdot} = \mathbf{e}_{a_1} - \mathbf{e}_{a_2}$ where $\mathbf{e}_i$ is the unitary vector with $1$ at index $i$. This matrix has rank $n-1$, but the constraint $\sum_{a \in \mathbb{A}} s_a = 0$ guarantees an identifiable model.

\subsubsection{Probit Models}

In some cases, especially for tractability in Bayesian models, we may prefer to replace the extreme-value distributed player performances with Gaussian player performances. In an $n$-player game, let
\begin{align}
   Y_{a_i} \overset{\text{ind}}{\sim} N(s_{a_i},1) \text{ for } i = 1,\dots,n. \label{normperf}
\end{align}

For instance, a 2-player game using model (\ref{2plperf}) yields
\begin{align}
    P_{a_1,a_2}(o) = \begin{cases}
      \Phi\left(\frac{s_{a_1} - s_{a_2}}{\sqrt{2}}\right) & \text{if}\ o=\text{(Win, Lose)} \\
      \Phi\left(\frac{s_{a_2} - s_{a_1}}{\sqrt{2}}\right) & \text{if}\ o=\text{(Lose, Win)},
    \end{cases} \label{prob}
\end{align}
where $\Phi(\cdot)$ is the cumulative distribution function of the standard normal distribution. Again, this model is a binary regression model as above but now with a probit link function. For simplicity, we proceed using probit links, but the principles discussed generalize to common link functions.

\subsubsection{Ties}

Extending our models to accommodate tied games will be sufficient for the vast majority of 2-player games. Introducing a tie threshold $t$ yields:
\begin{align}
O|Y_{a_1},Y_{a_2},t = \begin{cases}
(\text{Win, Lose}) & \text{if}\ Y_{a_1} - Y_{a_2}> t \\
(\text{Tie, Tie}) & \text{if}\ -t \leq Y_{a_1} - Y_{a_2} \leq t \\
(\text{Lose, Win}) &\text{if}\ \text{otherwise} \end{cases} \label{2pltie}
\end{align}
for some $t \in \mathbb{R}^+$. Using the latent performance model in (\ref{normperf}), the outcome probabilities are:
\begin{align}
    P^t_{a_1,a_2}(o) = \begin{cases}
      \Phi\left(\frac{s_{a_1} - s_{a_2} - t}{\sqrt{2}}\right) & \text{if}\ o=(\text{Win, Lose}) \\
      \Phi\left(\frac{t - (s_{a_2} - s_{a_1})}{\sqrt{2}}\right) - \Phi\left(\frac{-t-(s_{a_1} - s_{a_2})}{\sqrt{2}} \right) & \text{if}\ o=(\text{Tie, Tie})\\
      \Phi \left(\frac{ -t - (s_1 - s_2)}{\sqrt{2}}\right) &\text{if}\ o=(\text{Lose, Win}).
    \end{cases} \label{probt}
\end{align}

This model is no longer strictly a GLM but a ``cumulative link model'' \citep{christensen2013, ordinalprimer}. Nonetheless, \citet{burridge} has shown that in this case (\ref{mle}) is concave, so global maximization is straightforward. 

\subsubsection{Multiplayer Games}

For multiplayer games without ties, it would be natural to use a skill model equivalent to the multinomial generalized linear model \citep{agresti}. Thus
\begin{align}
    O|Y_{\mathbf{a}} = \text{``Player } a_j \text{ Wins''} \text{ if } Y_{a_j} \geq Y_{a_1},\dots,Y_{a_n}.
\end{align}
Again, using model \ref{normperf} results in
\begin{align}
    P_\mathbf{a}(o = \text{``Player } a_j \text{ Wins''}) &= \underset{k \neq j}{\prod_{k=1}^n} \Phi\left(\frac{s_{\mathbf{a}_j} - s_{\mathbf{a}_k}}{\sqrt{2}}\right).
\end{align}

\subsubsection{Latent Performance Based Luck Metrics}

The metric $\mathcal{L}$ is quite general, as it depends only on the assumption that there exist for each player fixed probabilities that they will achieve various outcomes for each combination of opponents. For example, in a two-player game without ties, $\mathcal{L}$ depends on all of the win probabilities $\{p_{ij}|i \neq j; i,j \in \mathbb{A}\}$, with only the constraint that $P_{ij} + P_{ji}$ = 1, thus having $\binom{A}{2}$ free parameters. In theory, we could estimate these probabilities entirely non-parametrically from enormous balanced data sets in which all combinations of players are represented, but generally we are compelled to use skill models. 

Contingent on these more constrained models, simpler metrics may have more intuitive force. In a unidimensional latent performance skill model, we may consider an alternate quantification of luck and returns to skill. Let $S = \mathbb{R}$, and $\sum_{a \in \mathbb{A}}s_a = 0$. Then let
\begin{align}
    \ell^2(\mathbf{s}) = \frac{1}{1 + \frac{1}{n}\sum_{a \in \mathbb{A}} s_a^2} = 1 - \text{\textcursive{s}}^2(\mathbf{s}).
\end{align}

Then $\ell^2(\mathbf{s}) \in [0,1]$ and can be interpreted as a measure of the fluctuation in player performance (fixed to 1) from game to game relative to the spread of player skill levels amongst the population. A further simplification is contingent on the assumption that the skill levels are normally distributed in the population:
\begin{align}
    \mathbf{s} \sim N(0,\sigma_s^2 \mathbb{I}_n).
    \label{prior}
\end{align}

In this model, the skill levels are latent variables, and as in a random-effects model, we may study the similar quantity
\begin{align}
    \ell^2(\sigma_s) = \frac{1}{1 + \sigma_s^2} = 1 - \text{\textcursive{s}}^2(\sigma_s)
\end{align}
which is an ``intra-player'' correlation.

Because $\ell^2(\mathbf{s})$ and $\ell^2(\sigma)$ depend only on variation in skill levels and performances, they do not depend on the specific nature of the game outcome (binary, ordinal, or multinomial with various numbers of categories). Therefore, they may be especially useful in comparing the amount of luck across games at different player counts.

\subsection{Fitting Algorithms}

The following algorithms can be used to estimate $\mathcal{L}$ and $\ell^2$ in two-player games with ties under different assumptions. The first, a frequentist algorithm, corresponds to a model with fixed skill levels, while the second, a Bayesian algorithm, corresponds to the hierarchical model in which the skills are normally distributed in the population. As shown in Section \ref{empiricalstudy}, the second algorithm is especially useful in the common case of highly unbalanced outcome data.

\subsubsection{Method 1: Skill Estimation}

Suppose there is a population of $A$ players, and a set of $n$ games observed. In game $i$, player $a_{1i}$ with skill level $s_{a_{1i}}$ competes against player $a_{2i}$ with skill level $s_{a_{2i}}$, $i \in \{1,\dots,A\}$. Let $o_i$ be the outcome of the game, where
\begin{align*}
    o_i = \begin{cases}
      1 & \text{if player $a_{1i}$ wins} \\
      0 & \text{if the players tie} \\
      -1 &\text{if player $a_{2i}$ wins}.
    \end{cases}
\end{align*}

Then using the model in \ref{2pltie}, the data likelihood to be maximized is
\begin{align*}
    l(\mathbf{o};\mathbf{a},t) &= \sum_{i=1}^n \log P(O_i = o_i). \\
\end{align*}

Let $x_{i \cdot} = \mathbf{e}_{a_1} - \mathbf{e}_{a_2}$ where $\mathbf{e}_i$ is the unitary $A$-dimensional vector with a 1 at index $i$. Then $\mathbf{x}$ is the $n \times A$ design matrix in this cumulative link model. Note that $\mathbf{x}$ is extremely sparse with only $2n$ nonzero entries. 

There are two obstacles to fitting the model using straightforward likelihood maximization. The first is that the matrix $\mathbf{x}$ is column-rank deficient. Consider a graph $G$ of the population of players, with edges connecting any two players who have played a least one game together. The rank of $\mathbf{x}$ is $A-K$ where $K$ is the number of separate subgraphs of $G$. 

The second obstacle is that in data sets containing data on some players who have played few games, there are likely to be some players who have only ever lost or only ever won. This is a case of perfect data separation, and the maximum likelihood estimate for the skill levels of those players does not exist. The algorithm will produce extremely large estimates for those players' skill levels, and consequently  the $\ell^2(\mathbf{s})$ for the game will be found to be extremely low. 

The first obstacle can be solved by including constraints in the model of the form $\sum_{a=1}^{A_k} s_{k_a} = 0$ for $k \in \{1,\dots,K\}$. Thus every subcommunity of players must have skill levels centered at 0. 

A second solution, which addresses both challenges, is to introduce a small ridge penalty which obviates the need for constraints and prevents oversized skill level estimates due to perfect separation. Using this strategy, the penalized loss function is
\begin{align*}
    l_\lambda(\mathbf{o};\mathbf{s},t) &= \sum_{i=1}^n \log P(O_i = o_i) - \lambda ||\mathbf{s}||_2^2\\
    &= \sum_{i=1}^n l_i - \lambda ||\mathbf{s}||_2^2
\end{align*}
where
\begin{align*}
    l_i = \begin{cases}
      \log(1-\Phi(\frac{t-x_{i\cdot}^\top \mathbf{s}}{\sqrt{2}})) & \text{if } o_i = 1 \\
      \log(\Phi(\frac{t-x_{i \cdot}^\top \mathbf{s}}{\sqrt{2}}) - \Phi(\frac{-t - x_{i \cdot} ^\top \mathbf{s}}{\sqrt{2}})) & \text{if } o_i = 0 \\
      \log(\Phi(\frac{-t-x_{i \cdot}^\top \mathbf{s}}{\sqrt{2}})) &\text{if } o_i = -1.
    \end{cases}
\end{align*}

The Newton-Raphson algorithm \citep{lange} for optimizing this loss function is:
\begin{align}
    \begin{bmatrix} \mathbf{s} \\ t \end{bmatrix}^{j+1} =   \begin{bmatrix} \mathbf{s} \\ t \end{bmatrix}^{j} - \left(\nabla^2(l_\lambda(\mathbf{o};\mathbf{s},t))\right)^{-1} \nabla(l_\lambda(\mathbf{o};\mathbf{s},t)). \label{NRalgo}
\end{align}
$\nabla^2(l_\lambda(\mathbf{o};\mathbf{s},t))$ and $\nabla(l_\lambda(\mathbf{o};\mathbf{s},t))$ are given in Appendix \ref{NRform}. 

\subsubsection{Method 2: Population Variance Estimation}
\label{popvar}

Modeling the population of skill levels and latent performances, we have
\begin{align*}
    \mathbf{s} &\sim N(0,\sigma_s^2 \mathbb{I}_A) \\
    y_i\mid \mathbf{s} &\overset{\text{iid}}{\sim} N(s_{a_{1i}}-s_{a_{2i}},2) \text{ for } i \in \{1,\dots,n\} \\
    o_i \mid y_i,t &= \begin{cases}
    1 &\text{if } y_i > t \\
    0 &\text{if } -t \leq y_i \leq t \\
    -1 &\text{otherwise}.
    \end{cases} 
\end{align*}

We could use a stochastic EM algorithm \citep{diebolt1996stochastic} to find maximum likelihood estimates for $t$ and $\sigma_s^2$, or use a fully Bayesian approach using prior distributions for these variables. Modeling $t$ conditional on $\mathbf{s}$ results in unacceptably slow mixing times, therefore it is useful to collapse the chain by modeling $t$ conditional only on $\sigma_s^2$ and $y$ as follows:
\begin{align*}
    \sigma_s^2 &\sim \Gamma^{-1}(a_\sigma,b_\sigma),\,\,
    p \sim \text{beta}(a_p,b_p)\, \,  {\text{and}} \,\, t = \sqrt{2(1+\sigma_s^2)}\Phi^{-1}(\frac{1+p}{2})
\end{align*}
where $\Gamma^{-1}(\cdot,\cdot)$ denotes the inverse-gamma distribution parametrized by shape and scale, and $\text{beta}(\cdot,\cdot)$ denotes the beta distribution. This allows us to sample $p$ conditional only on the outcomes, $\mathbf{o}$. The full conditional distributions which facilitate Gibbs sampling can be found in Appendix \ref{conditionals}. 

\section{Empirical Study}
\label{empiricalstudy}
\subsection{Professional Sports}

How much luck is in a single Major League Baseball (MLB), National Football League (NFL), National Hockey League (NHL) or National Basketball Association (NBA) game for the two competing teams? These games provide a great place to start because data on the outcomes is widely available (we used {\tt www.sports-reference.com}), but also because the skill levels of the teams can reasonably be assumed to be constant over the course of a season and the match schedules are well balanced. 

First, assuming the model in (\ref{probt}), using the regularized sparse cumulative link model as in Method 1 to estimate $\mathcal{L}$ and $\ell^2$ for each of the three games over three consecutive seasons (2015-2017), with the results ($\lambda = .3$) in Table 2.  This method works reasonably well for NHL, NBA and MLB data, as large number of balanced matches wash out the effect of the regularization parameter (see Figure 1). However, due to the small number of NFL matches played each year, the regularization parameter $\lambda$ largely determines $\mathcal{L}$ and $\ell^2$.

\begin{table}
\caption{$\mathcal{L}$ and $\ell^2$ values using penalized skill estimates with $\lambda = .3$} 
\centering
\begin{tabular}{c c|c|c|c|c}
    &  Season & Teams & Matches & $\hat{\mathcal{L}}$ & $\hat{\ell^2}$ \\\hline
    &  2015 & 32 & 256 & .908 & .647\\  
NFL &  2016 & 33 & 336 & .930 & .722\\
    &  2017 & 32 & 256 & .899 & .594\\
    \hline
    &  2015 & 30 & 1230 & .982 & .927\\
NHL &  2016 & 30 & 1230 & .976 & .901\\
    &  2017 & 31 & 1271 & .973 & .890\\
    \hline
    & 2015 & 30 & 1316 & .920 & .682\\
NBA & 2016 & 30 & 1309 &  .946 & .779\\
    & 2017 & 30 & 1312 & .939 & .761\\
    \hline 
    & 2015 & 30 & 2429 & .987 & .945 \\
MLB & 2016 & 30 & 2427 & .988 & .949
\end{tabular}
\end{table}

\begin{figure}[H]\label{figure1}
    \centering
    \includegraphics[scale=.75]{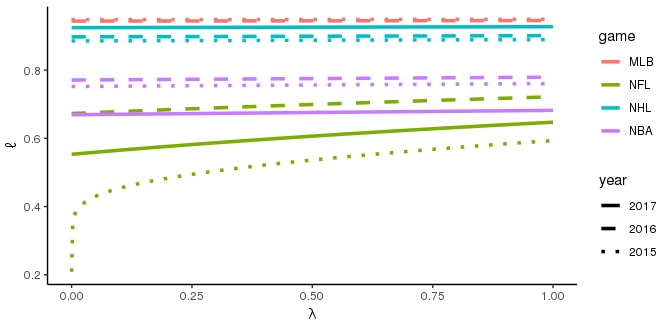}
    \caption{$\ell^2$ for varying values of $\lambda$.}
    \label{fig:sportslam}
\end{figure}

Method 2 provides a convenient Bayesian alternative with error quantification, the results of which are shown in Table \ref{sportsbayes}.

\begin{table}
\caption{\label{sportsbayes}Posterior estimate and standard deviation of $\ell^2$ using $a_\sigma = 2, b_\sigma = 1$. Thus the prior mean for $\sigma_s^2$ is $1$, corresponding to equal variation in outcomes due to skill and performance.} 
\centering
    \begin{tabular}{c c|c|c|c|c}
    &  Season & Teams & Matches &  E($\ell^2 \mid \cdot$) & sd($\ell^2 \mid \cdot$) \\
    \hline
    &  2015 & 32 & 256 & .694 & .065\\  
NFL &  2016 & 33 & 336 & .746 & .051\\
    &  2017 & 32 & 256 & .623 & .078\\
    \hline
    &  2015 & 30 & 1230 & .882 & .028\\
NHL &  2016 & 30 & 1230 & .866 & .031\\
    &  2017 & 31 & 1271 & .855 & .033\\
    \hline
    & 2015 & 30 & 1316 & .682 & .060\\
NBA & 2016 & 30 & 1309 &  .764 & .050\\
    & 2017 & 30 & 1312 & .742 & .051\\
    \hline 
    & 2015 & 30 & 2429 & .898 & .026 \\
MLB & 2016 & 30 & 2427 & .901 & .024
\end{tabular}
\end{table}

Due to the fact that the populations examined by these data sets consist of the most elite teams playing at the highest level, it would not be reasonable to characterize the hobbyist version of these sports using these luck metrics. Regardless of how skill intensive a game is, if we restrict our scope to only the best players, we will usually find that the outcomes are determined mostly by luck. \citet{success} refers to this phenomenon as the ``paradox of skill.'' Here the most we can say is that these elite teams are so similar in their high levels of skill that the outcomes of all of these games are determined mostly by happenstance. However, NFL teams manifest the most differences in effective skill, followed by the NBA, then the NHL, and finally, knowing which MLB teams are playing seems to give us very little information about who will win. These results roughly agree with those in \citet{success}, with NBA notably lower in luck than NHL and MLB. However, whereas we estimate the amount of luck inherent to each match, Mauboussin estimates the luck in an entire \textit{season} of each game, thus he finds NFL, with its short season, to be high in luck. We find assessing the single-match luck allows us to more naturally compare the amount of luck across sports with seasons of different lengths. 

\subsection{Chess and Go}

Go is sometimes considered a more opaque game than Chess and less friendly to beginners. At the professional level, does Chess have more luck than Go?

We use a chess data set from {\tt Kaggle.com} which contains 65,053 games played by the world's top 13,000 chess players over the last 12 years \citep{kagchess}.  For Go, we use a widely available set of 42,302 professional games played since 2000 \citep{godata}.

These data sets are less amenable to Method 1 than the sports data sets, in which each team plays a moderate number of games. In the Chess and Go data sets, a large proportion of players have played relatively few games. Thus these players' skill levels are largely determined by our choice of $\lambda$, and so is our estimate of $\mathcal{L}$ and $\ell^2$. Setting $\lambda = 0$, there is unresolved perfect separation in the data resulting in infinite skill estimates and the luck metrics found to be 0. 

One solution is to use cross validation to find the value of $\lambda$ which captures the most predictive spread in skill levels. However, optimal values of $\lambda$ from cross validation are sensitive to the number of folds in the cross-validation procedure, and furthermore this estimation procedure does not lead to straightforward error quantification.  Thus Method 2 is attractive here. Below are posterior estimates of $\ell^2$ and $t$ for Chess and Go using $a_\sigma = 2, b_\sigma=1, a_p = 2, b_p = 5$ with 250 samples after a burn-in period of 100 iterates. Note the game of Go does not allow for ties. On the other hand, professional Chess games in this data set results in a tie 44.1\% of the time!

\begin{table} \label{tab:chessgo}
\caption{Posterior summary statistics for $\ell^2(\sigma_s)$ and $t$ for Go and Chess.}
\centering
\begin{tabular}{l|c|c|c|c|c|c}
 & Players & Matches & $\text{E}(\ell^2 \mid \cdot)$ & $\text{sd}(\ell^2 \mid \cdot)$ & $\text{E}(t \mid \cdot)$ & $\text{sd}(t \mid \cdot)$ \\
 \hline
 Go & 2,312 & 42,162 & .586 & .014 & 0 & 0 \\
 Chess & 7,138 & 64,953 & .640 & .007 & 1.03 & .008 \\
\end{tabular}
\end{table}

Thus we conclude that in this population of professional Go players, there is somewhat more variation in effective skill amongst the players.  If we are willing to assume that these populations represent similar swaths of the highest level of play for each game, we can claim that at the highest levels, Chess has more luck than Go. 

\subsection{\textit{Race for the Galaxy} and \textit{Seasons}}

Moving on to the more niche boardgame world, we analyze data on two games graciously provided by {\tt boardgamearena.com}: \textit{Race for the Galaxy} and \textit{Seasons}. Among the reasons to suspect that \textit{Race for the Galaxy} has more luck are that it is shorter, actions are selected simultaneously rather than sequentially, and cards are drawn directly from the deck rather than drafted. 
The advantage of using data from an online gaming platform like {\tt boardgamearena.com} is that it may be reasonable to assume that the population of players playing each game is quite similar. However, because these players are not professionals, their skill levels likely change dramatically over the course of their matches. Because our algorithm assumes skill levels are fixed, we will overestimate the amount of luck in these games due to the fact that we will attribute changes in skill level to random fluctuation in performance. In this paper, we will abate this limitation by using data only for players who, at the time of the match, have played between 150 and 250 games. In that range, we can assume that player skill levels are increasing quite slowly. 
Below are posterior estimates of $\ell^2$ and $t$ for \textit{Race for the Galaxy} and \textit{Seasons} using $a_\sigma = 2, b_\sigma=1, a_p = 2, b_p = 5$ with 250 samples after a burn-in period of 100 iterates. 

\begin{figure}
    \centering
    \includegraphics[scale = .5]{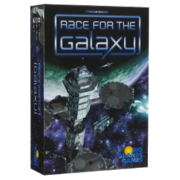} 
    \hspace{10mm}
    \includegraphics[scale = .5]{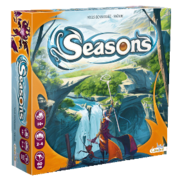}
    \caption{\textit{Race for the Galaxy} and \textit{Seasons} are both tableau-building non-collectable dedicated card games.}
    \label{fig:my_label}
\end{figure}

\begin{table}
\caption{Posterior summary statistics for $\ell^2(\sigma_s)$ and $t$ for \textit{Race for the Galaxy} and \textit{Seasons}}
\centering
\begin{tabular}{l|c|c|c|c|c|c}
 & Players & Matches & $\text{E}(\ell^2 \mid \cdot)$ & $\text{sd}(\ell^2 \mid \cdot)$ & $\text{E}(t \mid \cdot)$ & $\text{sd}(t \mid \cdot)$ \\
 \hline
 Race for the Galaxy & 4,265 & 21,978 & .998 & .00007 & .006 & .0007 \\
 Seasons & 7,138 & 64,953 & .818 & .014 & .003 & .0005
\end{tabular}
\end{table}

It is encouraging to see that both of these more casual boardgames have considerably more luck than famously ``pure-skill'' games like Chess and Go. As expected, \textit{Race for the Galaxy} has more luck than \textit{Seasons}; in fact, it seems that when \textit{Race for the Galaxy} players who have played between 150 and 250 games play against one another, the results are almost completely unpredictable. On the other hand, in \textit{Seasons}, certain players have discovered strategies which give them a substantial edge.  

\section{Conclusion}

This paper builds on the ludological foundations in \cite{cog}, formalizing the concepts of ``Returns to Skill'' and Luck. We propose a framework for thinking about luck which contextualizes the formulations of \cite{cog} and \cite{success} and generalizes the methodologies inspired by these formulations. Furthermore, we have proposed two metrics for luck in games: one information-theoretical, and one based on Gaussian models for player skill and performance. We have also proposed two statistical methods for estimating luck: a maximization algorithm which is fast but often stymied by real-life game data, and a sampling algorithm which is robust to pathological data distributions.  

Empirical evaluations of luck in games provides a grounding for theories about game mechanisms and populations of players. Although game outcome data alone does not allow us to distinguish between the amount of luck ``inherent to a game'' and the amount of luck manifest in a particular population, with further assumptions or data about the level of dedication or training of the players, we can make relevant quantitative statements about games using data about matches and their outcomes.

Luck is one of the defining characteristics of games, but we expect that other aspects of games can be meaningfully quantified. Are there games for which skill is best modelled as multidimensional? How can we characterize the degree of player interaction in a game? How do players' odds of winning evolve over the course of a game? For quantitative inquiries into intractable games, let's look to gameplay data for insights.

\newpage
\nocite{*}
\bibliographystyle{chicago}
\bibliography{references} 

\newpage

\appendix
\section{Newton-Raphson Updates for Regularized Skill Estimation}
\label{NRform}

For Algorithm \ref{NRalgo}, we have
\begin{align*}
    l_\lambda(\mathbf{o};\mathbf{s},t) &= \sum_{i=1}^n \log P(O_i = o_i) - \lambda ||\mathbf{s}||_2^2\\
    &= \sum_{i=1}^n l_i - \lambda ||\mathbf{s}||_2^2,
\end{align*}
where
\begin{align*}
    l_i = \begin{cases}
      \log(1-\Phi(\frac{t-x_{i\cdot}^\top \mathbf{s}}{\sqrt{2}})) & \text{if } o_i = 1 \\
      \log(\Phi(\frac{t-x_{i \cdot}^\top \mathbf{s}}{\sqrt{2}}) - \Phi(\frac{-t - x_{i \cdot} ^\top \mathbf{s}}{\sqrt{2}})) & \text{if } o_i = 0 \\
      \log(\Phi(\frac{-t-x_{i \cdot}^\top \mathbf{s}}{\sqrt{2}})) &\text{if } o_i = -1.
    \end{cases}
\end{align*}

The Newton-Raphson algorithm for optimizing this loss function is:
\begin{align*}
    \begin{bmatrix} \mathbf{s} \\ t \end{bmatrix}^{j+1} =   \begin{bmatrix} \mathbf{s} \\ t \end{bmatrix}^{j} - \left(\nabla^2(l_\lambda(\mathbf{o};\mathbf{s},t))\right)^{-1} \nabla(l_\lambda(\mathbf{o};\mathbf{s},t))
\end{align*}
where the above gradient and Hessian are given by
\begin{align*}
    \nabla(l_\lambda(\mathbf{o};\mathbf{s},t)) &= \sum_{i=1}^n \nabla_i^o - \lambda \mathbf{s}\\
    \nabla^2(l_\lambda(\mathbf{o};\mathbf{s},t)) &= \sum_{i=1}^n (\nabla_i^o)^2 - \lambda \mathbb{I}_A
\end{align*}
where
\begin{align*}
\nabla_i^1 &= S^1(\eta_i^1)\begin{bmatrix} x_{i \cdot}^\top \\ {-1} \end{bmatrix} \\
(\nabla_i^1)^2 &= (\eta_i^1 - S^1(\eta_i^1))S^1(\eta_i^1) \begin{bmatrix} x_{i \cdot} \\ {-1} \end{bmatrix} \begin{bmatrix} x_{i \cdot}^\top & {-1} \end{bmatrix}\\
\nabla_i^{-1} &= -S_2(\eta_i^2)\begin{bmatrix} x_{i \cdot} \\ 1 \end{bmatrix} \\
(\nabla_i^{-1})^2 &= -(\eta_i^2 + S^2(\eta_i^2))S^2(\eta_i^2) \begin{bmatrix} x_{i \cdot} \\ {1} \end{bmatrix} \begin{bmatrix} x_{i \cdot}^\top & {1} \end{bmatrix} \\
\nabla_i^0 &= \begin{bmatrix} -\left(\frac{\phi(\eta_i^1) - \phi(\eta_i^2)}{\Phi(\eta_i^1) - \Phi(\eta_i^2)} \right) x_{i \cdot} \\
\frac{\phi(\eta_i^1) + \phi(\eta_i^2)}{\Phi(\eta_i^1) - \Phi(\eta_i^2)}
\end{bmatrix} \\
(\nabla_i^0)^2 &= [\Phi(\eta_i^1) - \Phi(\eta_i^2)]^{-2}\begin{bmatrix} (1) & (2) \\ (2)^\top & (3) \end{bmatrix}
\end{align*}
where the matrix entries are
\begin{align*}
(1) &= \left(-[\eta_i^1 \phi(\eta_i^1) - \eta_i^2 \phi(\eta_i^2)][\Phi(\eta_i^1) - \Phi(\eta_i^2)] - [\phi(\eta_i^1) - \phi(\eta_i^2)]^2\right) x_{i \cdot} x_{i \cdot}  ^\top \\
(2) &= \left([\eta_i^1\phi(\eta_i^1) + \eta_i^2\phi(\eta_i^2)][\Phi(\eta_i^1) - \Phi(\eta_i^2)] + [\phi^2(\eta_i^1) - \phi^2(\eta_i^2)]\right) x_{i\cdot} \\
(3) &= \left(-[\eta_i^1 \phi(\eta_i^1) - \eta_i^2\phi(\eta_i^2)][\Phi(\eta_i^1) - \Phi(\eta_i^2)]-[\phi(\eta_i^1) + \phi(\eta_i^2)]^2 \right)
\end{align*}
and
\begin{align*}
    \eta_i^1 &= \frac{t - x_i^\top \mathbf{s}}{\sqrt{2}} \\
    \eta_i^2 &= \frac{-t - x_i^\top \mathbf{s}}{\sqrt{2}} \\
    S^1(\eta) &= \frac{\phi(\eta)}{1-\Phi(\eta)} \\
    S^2(\eta) &= \frac{\phi(\eta)}{\Phi(\eta)}.
\end{align*}

\section{Conditional Distributions for Gibbs Sampling}
\label{conditionals}
Using the distributional assumptions for Method 2 in Section \ref{popvar}, the full conditional distributions are as follows. We can sample the latent performances $\mathbf{y}$ and the skill levels $\mathbf{s}$ as blocks:
\begin{align*}
    \mathbf{y_i} \mid \mathbf{s},t \overset{\text{ind}}{\sim} N(s_{a_{1i}} - s_{a_{2i}},2)\mathbf{1}_{a_i \leq y_i \leq b_i} \text{ for } i \in {1,\dots,n}, \\
\end{align*}
where 
$$a_i,b_i = \begin{cases}
t, \infty &\text{if } o_i = 1 \\
-t, t &\text{if } o_i = 0 \\
-\infty, -t &\text{otherwise}
\end{cases}$$

\begin{align*}
    \mathbf{s} \mid \mathbf{y} &\sim N_A\left((\mathbf{x}^\top \mathbf{x} + \frac{\sqrt{2}}{\sigma_s^2})^{-1}x^\top \mathbf{y}, (\mathbf{x}^\top \mathbf{x} + \frac{\sqrt{2}}{\sigma_s^2})^{-1}\right),
\end{align*}
and
\begin{align*}
    \sigma_s^2 \mid \mathbf{s} &\sim \Gamma^{-1}(a_\sigma + \frac{A}{2},b_\sigma + \frac{||\mathbf{s}||_2^2}{2})
\end{align*}
and we collapse the chain by sampling $p$ directly conditional on $o$. Let $|\mathbf{o}_0|$ be the number of ties in the data set. Then $
    p \mid \mathbf{o} \sim \text{beta}(a_p + |\mathbf{o}_0|,b_p + n - |\mathbf{o}_0|)$ and  $t = \sqrt{2(1+\sigma_s^2)}\Phi^{-1}(\frac{1+p}{2})$.
\end{document}